\documentclass[11pt]{article}

\usepackage{amsmath}

\usepackage[cp1251]{inputenc}

\newcommand{\N}{N\raise.7ex\hbox{\underline{$\circ $}}$\;$}

\begin{document}

\begin{center}

{\bf  E.M. Ovsiyuk,   V.V.  Kisel,   V.M. Red'kov  \\[3mm]
On exact solutions of the  Dirac equation \\ in a homogeneous magnetic field  in the Lobachevsky space}\\[3mm]
{\em Institute of Physics, NAS of Belarus\\
Belarussian  State Pedagogical University}

\end{center}

\begin{quotation}

There are constructed exact solutions of the quantum-mechanical Dirac
equation for a spin S=1/2 particle in Riemannian space of constant negative curvature,
 hyperbolic Lobachevsky space,  in presence of an external magnetic field, analogue of
  the homogeneous magnetic field in the Minkowski  space. A generalized formula for energy levels,
   describing   quantization of the motion of the particle in magnetic field on the background of
   the Lobachevsky space geometry, has been obtained.

   \end{quotation}

{\bf  1.  Introduction }

The  quantization  of a quantum-mechanical particle in the
homogeneous magnetic field belongs to classical  problems in
physics  \cite{1,2,3}. In  \cite{4,5,6}, exact solutions for  a
scalar particle in extended problem, particle in  external
magnetic field on the background of Lobachevsky $H_{3}$  and
Riemann $S_{3}$ spatial geometries were found. A corresponding
system in the frames of classical mechanics was examibed in
 \cite{7,8,9}. In the present paper, we consider a similar problem for a particle with spin $1/2$ described
 by Dirac equation in Lobachevsky space in presence of the external magnetic field.

\vspace{2mm}

{\bf  2. Cylindric coordinates ant the Dirac equation in hyperbolic space $H_{3}$ }

\vspace{2mm}

 In the Lobachevsky space, let us use an extended cylindric coordinates
\begin{eqnarray}
dS^{2} =  dt^{2} -  \mbox{ ch}^{2} z ( d r^{2} + \mbox{sh}^{2} r
\; d \phi^{2} ) - dz^{2}\; ,
\nonumber\\
u^{1} = \mbox{ch} \; z \; \mbox{sh}\; r \cos \phi \; , \;\;
u^{2} = \mbox{ch} \; z \; \mbox{sh}\;r \sin \phi \; ,\;\;
u^{3} = \mbox{sh}\; z \; , \;\;  u^{0} = \mbox{cosh}\; z \;
\mbox{ch} \; r \; ; \label{A.1}
\end{eqnarray}

\noindent \noindent where  $x^{j}= (r,\; \phi,\; z )$: $ r \in [ 0
, + \infty )\; , \; \phi \in [ 0 , 2\pi ]\;, \; z \in (- \infty ,
= \infty ) \;$; the curvature radius $\rho$ is taken as a  unit of
the length. An analogue of usual homogeneous magnetic field is
defined as
\begin{eqnarray}
 A_{\phi} = -2B \; \mbox{sh}^{2} {r \over 2} = - B\; (\mbox{cosh}\; r -1 )\; .
\label{A.2}
\end{eqnarray}

\noindent To coordinates  (\ref{A.1})  there corresponds the tetrad
\begin{eqnarray}
 e_{(a)}^{\beta} = \left |
\begin{array}{llll}
1 & 0 & 0 & 0 \\
0 & \mbox{cosh}\;^{-1}z & 0 & 0 \\
0 & 0 & \mbox{cosh}\;^{-1}z\;\mbox{sinh}\;^{-1} r & 0 \\
0 & 0 & 0 & 1
\end{array} \right | \;  .
\label{A.3}
\end{eqnarray}

\noindent  Christoffel symbols  $\Gamma^{r}_{\;\;jk }$ and Rici rotation coefficients
 $\gamma_{abc}$  are
\begin{eqnarray}
\Gamma^{r}_{\;\;jk } = \left | \begin{array}{ccc}
0 & 0 & \mbox{th}\;z \\
0 & - \mbox{sh}\; r \; \mbox{cosh}\; r & 0 \\
\mbox{th}\;z & 0 & 0
\end{array} \right | \; , \qquad \qquad
\Gamma^{\phi}_{\;\;jk } = \left | \begin{array}{ccc}
0 & \mbox{cth}\; r & 0\\
\mbox{cth}\; r & 0 & \mbox{th}\; z \\
0 & \mbox{th}\; z & 0
\end{array} \right | \; ,
\nonumber\\
\Gamma^{z}_{\;\;jk } = \left | \begin{array}{ccc}
-\mbox{cosh}\; z \;\mbox{sh}\; z & 0 & 0\\
0 & -\mbox{sh}\; z \; \mbox{cosh}\; z \; \mbox{sh}^{2} r & 0 \\
0 & 0 & 0
\end{array} \right | \; ,\;
\nonumber
\end{eqnarray}
\begin{eqnarray}
 \gamma_{12 2} =
 { 1 \over \mbox{cosh}\; z \mbox{tanh}\; r} \; , \qquad
 \gamma_{31 1} =
 \mbox{tanh}\; z\; , \qquad  \gamma_{32 2} =
 \mbox{tanh}\; z\; .
 \nonumber
 \label{A.4}
\end{eqnarray}

\noindent A general covariant Dirac equation (for more detail see \cite{10})  takes the form
\begin{eqnarray}
\left [  \; i \gamma^{0} \partial_{t}  + {i \gamma^{1}  \over
\mbox {ch}\, z}
 (    \partial_{r} + {1 \over 2}  {1 \over \mbox {th}\, r}  )
 + \gamma^{2}  {    i   \partial_{\phi} + e  B (\mbox {ch}\,r -1)    \over \mbox {ch}\, z\; \mbox {sh}\, r}    +
 i \gamma^{3} ( \partial_{z} + \mbox {th}\, z   ) - M \right  ]  \Psi = 0\; .
\label{A.6}
\end{eqnarray}

\noindent With the  substitution
$\Psi = \varphi /  \sqrt{\mbox {sh}\, r}\,\mbox {ch}\,z$ eq. (\ref{A.6}) becomes simpler
\begin{eqnarray}
\left [  \;  i \gamma^{1}  {\partial \over \partial r }  +
\gamma^{2} {    i   \partial_{\phi} + e  B (\mbox {ch}\,r -1)
\over  \mbox {sh}\, r}   +
  \mbox {ch}\, z\; \left  (i  \gamma^{0} {\partial  \over  \partial t}  +
  i  \gamma^{3} { \partial \over  \partial z} -  M  \right  )  \right ] \varphi =0\,.
\label{A.7}
\end{eqnarray}

\noindent Solutions of this equation will be searched in the form
\begin{eqnarray}
\varphi = e^{-i\epsilon t} e^{im \phi} \left | \begin{array}{c}
f_{1}(r,z)\\ f_{2}(r,z)\\
f_{3}(r,z)\\
f_{4}(r,z) \
\end{array} \right | ,
\nonumber
\end{eqnarray}

\noindent so that
\begin{eqnarray}
\left [  i \gamma^{1}  {\partial \over \partial r }  -  \gamma^{2}
\;
 {m -  e B (\mbox {ch}\,  r -1)  \over \mbox {sh}\,  r }   +
  \mbox {ch}\, z\; \left  ( \epsilon   \gamma^{0}   +
  i  \gamma^{3} { \partial \over  \partial z} -  M  \right  )  \right ]  \left | \begin{array}{c}
f_{1}(r,z)\\ f_{2}(r,z)\\
f_{3}(r,z)\\
f_{4}(r,z) \
\end{array} \right |
 =0 \; .
 \label{A.8}
\end{eqnarray}

\noindent  Taking the  Dirac matrices in spinor basis,
we get radial equations for  $f_{a}(t,z)$
\begin{eqnarray}
  ( {\partial \over \partial r }  +   \mu  ) \; f_{4}
   +  \mbox {ch}  z  \;  { \partial f_{3} \over  \partial z}
+  i \; \mbox {ch} \; z    \; ( \epsilon   f_{3}   -  M  f_{1} )
=0\, ,
\nonumber\\
( {\partial \over \partial r }  -  \mu  ) \;f_{3} -  \mbox {ch}
\; z\;   { \partial f_{4} \over  \partial z}
 +   i  \; \mbox {ch} \; z  \; (  \epsilon  f_{4}   - M  f_{2} ) =0\, ,
 \nonumber\\
 ( {\partial \over \partial r }  +    \mu )  \; f_{2}
+ \mbox {ch}  \; z\;  { \partial f_{1} \over  \partial z}
 - i   \; \mbox {ch} \; z  \; ( \epsilon f_{1} -M f_{3} ) =0\, ,
 \nonumber\\
  ( {\partial \over \partial r }  -    \mu  ) \; f_{1}
-  \mbox {ch}   \;z \;  { \partial f_{2} \over  \partial z} - i
\; \mbox {ch} \; z  \; (   \epsilon f_{2} -M f_{4} ) =0\,  ,
 \label{A.9}
\end{eqnarray}

\noindent where
$ \mu (r) = [m -  e B (\mbox {ch}\,  r -1) ]/ \;  \mbox{sh}\,  r $.
With linear restriction
\begin{eqnarray}
f_{3} = A f_{1} ,  \qquad f_{4} = A f_{2}
 \label{A.10}
\end{eqnarray}

\noindent eqs. (\ref{A.9}) give
\begin{eqnarray}
  ( {\partial \over \partial r }  + \; \mu  ) \; f_{2}
   +   \mbox {ch}\, z  \;  { \partial f_{1} \over  \partial z}
+  i \; \mbox {ch}\, z  \; (  \epsilon      -  {M \over A}   ) \;
f_{1} =0\,,
  \nonumber\\
 ( {\partial \over \partial r }  +   \; \mu )  \; f_{2}
+ \mbox {ch}\, z \;  { \partial f_{1} \over  \partial z}
 + i   \; \mbox {ch}\, z  \; (  - \epsilon  +  M  A  )  \; f_{1} =0\,,
  \nonumber\\
 ( {\partial \over \partial r }  - \; \mu  ) \;f_{1}
-  \mbox {ch}\, z  \;   { \partial f_{2} \over  \partial z}
 +   i  \; \mbox {ch}\, z \; (  \epsilon     - {M \over A}    ) \; f_{2}  =0\,,
  \nonumber\\
  ( {\partial \over \partial r }  -  \; \mu  ) \; f_{1}
-  \mbox {ch}\, z  \; { \partial f_{2} \over  \partial z} + i  \;
\mbox {ch}\, z \; (  -  \epsilon  + M  A ) \; f_{2} =0\, .
\label{A.11}
\end{eqnarray}

\noindent The system (\ref{A.11})  is self-consistent only if
\begin{eqnarray}
\epsilon -{M \over A} = - \epsilon + M A \qquad  \Longrightarrow
\qquad A = A_{1,2}= {\epsilon \pm p \over M} \; , \;\; p =
\sqrt{\epsilon^{2} - M^{2}} \; . \label{A.12}
\end{eqnarray}

\noindent
So, the problem is reduced to the following  systems
\begin{eqnarray}
 ( {\partial \over \partial r }  +   \; \mu )  \; f_{2}
+ \mbox {ch}\, z \;  { \partial f_{1} \over  \partial z}
 + i   \; \mbox {ch}\, z  \; ( - \epsilon  + M  A  ) \;  f_{1} =0\,,
  \nonumber\\
  ( {\partial \over \partial r }  -  \; \mu  ) \; f_{1}
-  \mbox {ch}\, z  \; { \partial f_{2} \over  \partial z} + i  \;
\mbox {ch}\, z \; (  -  \epsilon   + M  A ) \;  f_{2} =0\, .
\label{A.13}
\end{eqnarray}

\noindent Thus, we have two different  (but similar) cases

\vspace{4mm}
 $
 AM =   \epsilon + p \; ,
 $
\begin{eqnarray}
 ( {\partial \over \partial r }  +   \; \mu )  \; f_{2}
+ \mbox {ch}\, z \;  ( { \partial  \over  \partial z}
 +  i  p   \; )\;   f_{1} =0\,,
 \nonumber\\
  ( {\partial \over \partial r }  -  \; \mu  ) \; f_{1}
-  \mbox {ch}\, z  \; ({ \partial  \over  \partial z} - i  \; p\;
)\; f_{2} =0\, ; \label{A.14}
\end{eqnarray}

$
 AM =   \epsilon - p  \; ,
$
\begin{eqnarray}
 ( {\partial \over \partial r }  +   \; \mu )  \; f_{2}
+ \mbox {ch}\, z \;  ( { \partial  \over  \partial z}
 -  i  p\;)   \;    f_{1} =0\,,
  \nonumber\\
  ( {\partial \over \partial r }  -  \; \mu  ) \; f_{1}
-  \mbox {ch}\, z  \; ( { \partial  \over  \partial z} + i  \; p\;
) \; f_{2} =0\, . \label{A.15}
\end{eqnarray}

For definiteness, let us consider the system   (\ref{A.14})
(transition to the case
 (\ref{A.15}) is performed by the formal change $p \Longrightarrow -p$).
 Let us search solutions in the form
\begin{eqnarray}
f_{1} = Z_{1} (z) \; R_{1} (r) \; ,  \qquad  f_{2} = Z_{2}(z) \;
R_{2} (r) \; . \label{A.16}
\end{eqnarray}

\noindent Eqs.   (\ref{A.14})  result in
\begin{eqnarray}
 ( {\partial \over \partial r }  +   \; \mu )  \; Z_{2} R_{2}
+ \mbox {ch}\, z \;  ( { \partial  \over  \partial z}
 +  i  p   \; )\;   Z_{1}  R_{1} =0\,,
\nonumber\\
  ( {\partial \over \partial r }  -  \; \mu  ) \; Z_{1} R_{1}
-  \mbox {ch}\, z  \; ({ \partial  \over  \partial z} - i  \; p\;
)\; Z_{2}  R_{2} =0\, . \label{A.17}
\end{eqnarray}

\noindent Introducing the separating constant  $\lambda$:
\begin{eqnarray}
 \mbox {ch}\, z  \; ({ \partial  \over  \partial z} + i  \; p\; )\;
Z_{1}  = \lambda  \; Z_{2} \; , \qquad \mbox {ch}\, z  \; ({
\partial  \over  \partial z} - i  \; p\; )\; Z_{2} = \lambda \;
Z_{1}  \;  \label{A.18}
\end{eqnarray}

\noindent we arrive at the radial system
\begin{eqnarray} ( {\partial \over \partial r }  +   \; \mu )  \;  R_{2} +
\lambda  \; R_{1}  =0\,, \qquad
   ( {\partial \over \partial r }  -  \; \mu  ) \;  R_{1}
-  \lambda   \; R_{2}  =0\, . \label{A.19}
\end{eqnarray}

{\bf 3. Solution of the equation in  $z$-variable}

\vspace{3mm}

From (\ref{A.18}) it follows a second  order differential equation for
 $Z_{1}(z)$:
\begin{eqnarray}
{d^{2}Z_{1} \over dz} + {\mbox {sh}\, z  \over \mbox {ch}\, z}
 {dZ_{1}\over dz} +  \left(   p^{2}+ip{\mbox {sh}\, z\over \mbox {ch}\,
z} -  {\lambda^{2}\over \mbox {ch}\,^{2}z}
 \right)Z_{1}=0 \, .
\label{A.20}
\end{eqnarray}

\noindent In a new variable  $  y = (1 + \mbox{th}\; z)/2$, eq.  (\ref{A.20})  will give
\begin{eqnarray}
\left [ 4y (1-y)
 {d \over d y} +
2  (1-2y) {d \over d y}  +
     p^{2} ( {1 \over 1- y}  + {1 \over y} ) +ip (  {1 \over 1-y } - {1 \over y}  )
-  4 \lambda^{2}
 \right ] \; Z_{1}=0 \, .
\label{A.21}
\end{eqnarray}

\noindent With the substitution  $Z_{1} = y^{A}
(1-y)^{B} Z (y)$, eq. (\ref{A.21}) leads to
\begin{eqnarray}
4y\,(1-y)\,{d^{2}Z\over dy^{2}}+4\left[2A\,+\,{1\over
2}-(2A\,+\,2B\,+1)\,y\right]{dZ\over dy}+
\nonumber \\
+\left[{2A\,(2A-1)+p\,(p-i)\over y}+{2B\,(2B-1)+p\,(p+i)\over
1-y}-4\,(A+B)^{2}-4\lambda^{2}\right]Z=0\,. \nonumber
\end{eqnarray}

\noindent Requiring
\begin{eqnarray}
 A= -{ ip \over 2} \; ,\; {1 + ip \over 2} \, ,  \qquad B = {ip\over
2}\, ,\;{1 -ip \over 2} \, ; \label{A.23}
\end{eqnarray}

\noindent the equation for  $Z_{1}$ is reduced to that of hypergeometric   type
\begin{eqnarray}
y\,(1-y)\,{d^{2}Z\over dz^{2}}+ \left[2A\,+\,{1\over
2}-(2A\,+\,2B\,+1)\,y\right]{dZ\over
dz}-\left[(A+B)^{2}+\lambda^{2}\right]Z=0\,  \label{A.24}
\end{eqnarray}

\noindent with parameters given by
\begin{eqnarray}
\alpha = + i\lambda + A + B \;, \qquad \beta = -i\lambda+A+B \;,
\qquad \gamma = 2A + {1\over 2} \; ,
 \nonumber
 \\
Z_{1} =  \left ( {e^{z} \over  \mbox{cosh}\;  z } \right )^{A} \left
( {e^{-z} \over  \mbox{cosh}\;  z } \right )^{B}  \; F(\alpha, \;
\beta, \; \gamma;\;  {e^{z} \over  2\; \mbox{cosh}\;  z } )  \; .
\label{A.25}
\end{eqnarray}

There arise four   possibility depending on the choice of  $A,B$ in
(\ref{A.23}), they provide us with solutions of different behavior
in the  region $z \rightarrow \pm \infty$.
\begin{eqnarray}
\underline{\mbox{Variant }\;\; 1}, \qquad A = {1 +ip \over 2} \;,
\; B = {1 - i p \over 2} \;, \qquad A+B = 1 \;, \; A - B =ip  \; ,
\nonumber
\\
 \alpha = i\lambda + 1   \; , \;  \beta =  -i\lambda + 1 \; ,   \;    \gamma = ip  + {3\over 2} \; ,
\qquad Z_{1} = {e^{ipz}  \over  \mbox{cosh}\;  z  } \; F(\alpha, \;
\beta, \; \gamma;\;  {e^{z} \over  2\; \mbox{cosh}\;  z } )  \; ,
 \nonumber
\\
z \rightarrow  + \infty \;  , \;\;  Z_{1}  \rightarrow {e^{ipz}
\over  e^{z}}  = 0\; , \qquad z \rightarrow  - \infty \;  , \;\;
Z_{1}  \rightarrow {e^{ipz} \over  e^{-z}}  = 0 \; . \label{A.27}
\end{eqnarray}
\begin{eqnarray}
\underline{\mbox{Variant }\;\; 2},
 \qquad  \qquad A = -{ip \over 2} \;, \; B = {ip \over 2} \;, \qquad A+B =  0 \;, \; A- B = -ip  \; ,
\nonumber
\\
 \alpha = i\lambda \;   , \;  \beta =   -i\lambda   \; , \;    \gamma = -ip + {1\over 2}
\qquad Z_{1} = e^{-ipz}   \; F(\alpha, \;  \beta, \; \gamma;\;
{e^{z} \over  2\; \mbox{cosh}\;  z } )  \; . \label{A.28}
\end{eqnarray}
\begin{eqnarray}
\underline{\mbox{Variant }\;\; 3}, \qquad A = {1 +i p \over 2} \;,
\; B = {i p \over 2} \;, \qquad  A+B = i p +1/2\;, \; A - B =1/2
\; , \nonumber
\\
 \alpha = + i\lambda + {1 \over 2} + i p   , \qquad   \beta =  -i\lambda +{1\over 2} + i p   \; , \qquad
     \gamma = ip +  {3\over 2} \; ,
\nonumber
\\
Z_{1} = {e^{z/2}  \over ( \mbox{cosh}\;  z  ) ^{ip+1/2} } \;
F(\alpha, \;  \beta, \; \gamma;\;  {e^{z} \over  2\; \mbox{cosh}\; z
} )  \; , \nonumber
\\
z \rightarrow  + \infty \;  , \;\;  Z_{1} \rightarrow e^{-ip  \; z
}  \; ,\qquad z \rightarrow  - \infty \;  ,  \;\;  Z_{1}
\rightarrow e^{+ip\; z  } e^{-\infty }  = 0  \; . \label{A.26}
\end{eqnarray}
\begin{eqnarray}
\underline{\mbox{Variant }\;\; 4},
  \qquad A =  -{i p \over 2} \;, \; B = {1 - ip \over 2} \;, \qquad A+B = -ip +1/2  \;, \; A - B = -1/2  \; ,
\nonumber
\\
 \alpha = + i\lambda +{1 \over 2}   -ip \;   , \qquad  \beta =   - i\lambda +{1 \over 2}  -ip  \; , \qquad     \gamma = -ip + {1\over 2}\; ,
\nonumber
\\
Z_{1} = {e^{-z/2}  \over  ( \mbox{ ch}\; z )^{-ip +1/2} } \;
 F(\alpha, \;  \beta, \; \gamma;\;  {e^{z} \over  2\; \mbox{cosh}\;  z } )  \; , \hspace{20mm}
\nonumber
\\
z \rightarrow  + \infty \;  , \;\;  Z_{1} \rightarrow e^{+ip  \; z
}  e^{-z} = 0 \; ,\qquad z \rightarrow  - \infty \;  ,  \;\; Z_{1}
\rightarrow e^{-ip\; z  }  \; . \label{A.29}
\end{eqnarray}

\vspace{5mm}

{\bf 4. Solution of the equations in  $r$-variable}

\vspace{5mm}

From radial equations   (\ref{A.19}) it follows a second order  equation for $R_{1}$:
\begin{eqnarray}
\left ( {d^{2} \over dr^{2} } - {d \mu \over dr}  - \mu^{2} +
\lambda^{2}  \right ) R_{1} = 0 \; . \label{A.30}
\end{eqnarray}

\noindent Remembering on the meaning of  $\mu (r)$ (for shortness let ua us note  $eB$  as $B$)
we obtain  explicit form of the equation for $R_{1}$:
\begin{eqnarray}
{d^{2}R_{1}\over dr^{2}}+\left[{m\,\mbox {ch}\,  r+B\,(\mbox
{ch}\,  r-1)\over \mbox {sh}^{2}  r }-{[m-B\,(\mbox {ch}\,
r-1)]^{2}\over \mbox {sh}^{2}  r }+\lambda^{2}\right]R_{1}=0\,.
 \label{A.26}
\end{eqnarray}

\noindent With the variable  $ y= (1+\mbox {ch}\, r ) /2 $.
 eq.   (\ref{A.26})  gives
 \begin{eqnarray}
y(1-y){d^{2}R_{1}\over dy^{2}}+\left({1\over
2}-y\right){dR_{1}\over dy}- \hspace{20mm} \nonumber
\\
-\left[\lambda^{2}+{m^{2}\over 4} \left({1\over y}+{1\over
1-y}\right)+{m\over 4} \left({1\over y}-{1\over 1-y}\right)+{mB
\over y}-B^{2} \left(1-{1\over y}\right)+{B\over
2y}\right]R_{1}=0\,. \label{A.27}
\end{eqnarray}

\noindent
Making the substitution
$ R_{1} = y^{A} (1-y)^{C} R (y) $, eq.  (\ref{A.27}) is reduced to
\begin{eqnarray}
y(1-y){d^{2}R\over dy^{2}}+ \left [  2A + {1\over 2}-(2A+2C +1)\;
y  \right] \,{dR\over dy} \; + \nonumber
\\
+ \left[ \; { A^{2}-A/2-m^{2}/4-m/4- mB - B^{2}-B/2\over y} \;
+\right. \nonumber
\\
\left.+ \; {C^{2}-C/2- m^{2}/4 + m/4\over 1-y}
-(A+C)^{2}-\lambda^{2}+B^{2} \; \right]R=0\,. \label{A.28}
\end{eqnarray}

\noindent Requiring
\begin{eqnarray}
 A= - {2B + m\over 2}\; ,\;  {2B + m +1 \over 2} \, , \qquad
 C = {m\over 2} ,\; {1 -m \over 2} \, . \label{A.29}
\end{eqnarray}

\noindent we arrive at an equation of hypergeometric type
\begin{eqnarray}
y(1-y){d^{2}R\over dy^{2}}+\left [ 2A+{1\over 2}-(2A+2C+1)\;
y\right ] \, {dR\over dy} - \left [ \; (A+C)^{2}+\lambda^{2}-B^{2}
\; \right ]R=0\,, \nonumber \label{A.30}
\end{eqnarray}

\noindent so that
\begin{eqnarray}
\alpha = A+C +\sqrt{B^{2}-\lambda^{2}}\;, \qquad \beta =
A+C-\sqrt{B^{2}-\lambda^{2}} \;, \qquad \gamma = 2A\,+\,{1\over 2}
\; , \nonumber
\\
R_{1} = ( 1 + \mbox{cosh}\; r) ^{A} \; (1 - \mbox{cosh}\; r)^{C} \;
F(\alpha, \beta, \gamma ;\; {1 +  \mbox{cosh}\; r \over 2} ) \; .
 \label{A.32}
\end{eqnarray}

To have wave solutions finite in the origin $r=0$ (corresponding geometrical points belong to
the axis  $z$: $ u_{0} = \mbox{cosh}\; z  , \;  u_{3} =
\mbox{sh}\; z  , \; u_{1}=0 , \; u_{2}=0$ ) and in infinity  $r
\rightarrow \infty$, we must take positive  $C$ and negative   $A$, such that  $C+A<0$:

\begin{eqnarray}
R_{1} = ( 1 + \mbox{cosh}\; r) ^{A} \; (1 - \mbox{cosh}\; r)^{C} \;
F(\alpha, \beta, \gamma ;\; {1 +  \mbox{cosh}\; r \over 2} ) \; ;
\qquad C \geq 0 \; , \;  \; A < 0 \; .
 \label{A.33}
\end{eqnarray}

Let us follows all four possibilities to choose the parameters
\begin{eqnarray}
1. \qquad C = {m \over 2}  \geq 0\;  , \qquad A = {2B + m +1 \over
2} < 0 \; ,\qquad  C+A = B + m + {1 \over 2}  < 0 \; ;
 \nonumber
\\
2. \qquad C = {1-m \over 2}  \geq 0 \; , \qquad A = {2B + m +1
\over 2} < 0 \; ,\qquad  C+A = B + 1 < 0 \; ;
\nonumber
\\
3.  \qquad \qquad \qquad C = {m \over 2}  \geq 0\;  , \qquad A =
- {2B + m  \over 2} < 0 \; ,\qquad  C+A = - B   < 0 \; ;
\nonumber
\\
4. \qquad C = {1-m \over 2}  \geq 0 , \qquad A =  - {2B + m  \over
2} < 0 \; ,\qquad  C+A = - B  - m  +{1 \over 2}  < 0 \; .
\nonumber
\end{eqnarray}

Therefore, only two variant are appropriate
\begin{eqnarray}
 3. \qquad   m > 0\; ; \qquad \qquad 4. \qquad   -2B < m \leq 1 \; .
 \label{A.33}
\end{eqnarray}

\noindent Respective expressions  for radial functions are
\begin{eqnarray}
\underline{ \mbox{Variant  } \; 3 \; ,  \qquad  m > 0 } \; ,
\qquad \qquad C = m / 2 \;  , \qquad A =  - B - m  / 2  < 0 \; ,
\nonumber
\\
R_{1} = ( 1 + \mbox{cosh}\; r) ^{-B - m/2} \; (1 - \mbox{cosh}\;
r)^{m/2} \;  F(\alpha, \beta, \gamma ;\; {1 +  \mbox{cosh}\; r \over
2} ) \; , \nonumber
\\
\alpha = -B +\sqrt{B^{2}-\lambda^{2}}\;, \;\;  \beta = -B
-\sqrt{B^{2}-\lambda^{2}} \;, \;\; \gamma = -2B - m + {1\over 2}
\; ;
 \label{A.34}
\end{eqnarray}

\noindent with the quantization rule
\begin{eqnarray}
\alpha = -n \; \; \Longrightarrow \;\;  \qquad
\sqrt{B^{2}-\lambda^{2}} = B - n \qquad \Longrightarrow \qquad
\lambda^{2} = + 2Bn - n^{2} \; .
 \label{A.35}
\end{eqnarray}

\noindent To have radial function finite at the infinity
 $r \rightarrow \infty $,  the following inequality  must be imposed
\begin{eqnarray}
A+ C + n < 0 \qquad \Longrightarrow \qquad    B -n > 0\; ;
 \label{A.36}
\end{eqnarray}

\noindent  which insures the positive square root  $\sqrt{B^{2}-\lambda^{2}} $ in
(\ref{A.35}).

\begin{eqnarray}
\underline{ \mbox{Variant  } \; 4 \; ,  \qquad  \qquad   -2B < m
\leq 1 } \; ,  \qquad C = 1/2 - m / 2 \;  , \nonumber
\\
 A =  - B - m  / 2  < 0 \; , \qquad
C+A = - B  - m  + 1 / 2  <  0 \; , \nonumber
\\
R_{1} = ( 1 + \mbox{cosh}\; r) ^{-B - m/2} \; (1 - \mbox{cosh}\;
r)^{1/2 - m/2} \;  F(\alpha, \beta, \gamma ;\; {1 +  \mbox{cosh}\; r
\over 2} ) \; , \nonumber
\\
\alpha = (- B  - m  + 1 / 2)
 +\sqrt{B^{2}-\lambda^{2}}\;,
 \nonumber
 \\
   \beta = (- B  - m  + 1 / 2 ) -\sqrt{B^{2}-\lambda^{2}} \;, \;\;
\qquad \gamma = -2B - m + {1\over 2} \; .
 \label{A.37}
\end{eqnarray}

\noindent The quantization rule is
\begin{eqnarray}
\alpha = - n \qquad \Longrightarrow \qquad
 \sqrt{B^{2}-\lambda^{2}} =   B  + m  - 1 / 2 - n
 \nonumber
 \\
\Longrightarrow \qquad \lambda^{2} = + 2B (n -m +{1\over 2})  - (n
-m +{1\over 2}) ^{2} \; .
 \label{A.38}
\end{eqnarray}

\noindent The inequality must be fulfilled
\begin{eqnarray}
A+ C + n < 0 \qquad \Longrightarrow \qquad   B  + m  - 1 / 2  - n
> 0\; , \label{A.39}
\end{eqnarray}

\noindent  which make positive  the  root $\sqrt{B^{2}-\lambda^{2}} $ in  (\ref{A.38}).

Thus, the energy spectrum for spin $1/2$ particle in the   magnetic  field
in the Lobachevsky space model is given by two formulae
\begin{eqnarray}
3. \qquad
  \qquad  n < B \;, \qquad  m>0 \; , \qquad
\lambda^{2} = + 2Bn - n^{2} \; ; \nonumber
\\[3mm]
4. \qquad \qquad   n < B  + m  - 1 / 2 \; , \qquad   -2B < m \leq
1    \; , \nonumber
\\
\lambda^{2} = + 2B (n -m +{1\over 2})  - (n -m +{1\over 2}) ^{2}\;
. \label{A.40}
\end{eqnarray}

The transition to the limit os  the flat Minkowski space
is realized in accordance with the rules
\begin{eqnarray}
\lambda ^{2} \rightarrow {P ^{2}_{z} \rho^{2}  \over \hbar^{2}  }
= \lambda_{0}^{2} \; \rho^{2} \; ,\qquad B \rightarrow {e B \over
\hbar c}\;  \rho^{2} \nonumber
\\
2. \qquad \lambda^{2}_{0}  =   {2e B \over \hbar } \; n   \; ;
\qquad\qquad 4. \qquad \lambda^{2}_{0}  =   {2e B \over \hbar }\;
(n -m +{1\over 2})   \; . \label{A.41}
\end{eqnarray}

In the end there should be given a  clarifying remarks.
In fact, the above used  relationship
$-i \partial_{\phi } \Psi = m  \; \Psi $ represents transformed from cartesian coordinates to cylindrical
an eigen-value  equation for the third projection of the the total angular momentum of the Dirac particle
\begin{eqnarray}
\hat{J}_{3} \Psi_{Cart}  = ( -i {\partial \over  \partial \phi }
  + \Sigma_{3} )\; \Psi_{Cart} = m  \; \Psi = m\; \Psi_{Cart} \; ;
\label{3.45}
\end{eqnarray}

\noindent  this means that for  the quantum number
 $m$ are permitted half-integer values $ m = \pm {1 \over 2}, \; \pm {3\over 2}, ... $


Authors are  grateful to participants of the Scientific Seminar of the Laboratory
of theoretical physics, Institute of physics of National Academy od Sciences of Belarus, for discussion and advices.


\begin{thebibliography}{xxx}




\bibitem{1} Rabi I.I.
 Das freie Electron  in Homogenen Magnetfeld nach der  Diraschen Theorie.
// Z. Phys.    {\bf 49},  507 -- 511 (1928).



\bibitem{2}  Landau  L., Diamagnetismus der Metalle,
 Ztshr. Phys.    {\bf 64},   629--637 (1930).




\bibitem{3}
 Plesset M.S.
 Relativistic  wave  mechanics of the electron  deflected  by magnetic field.
// Phys.Rev.   {\bf  12}, 1728 -- 1731 (1931).




\bibitem{4}
 Bogush A.A., Red'kov V.M., Krylov G.G..
Schr\"{o}dinger particle in magnetic and electric fields in
Lobachevsky and Riemann spaces. // Nonlinear Phenomena in Complex
Systems.  {\bf  11},  no 4,  403 -- 416 (2008).


\bibitem{5}
A.A. Bogush,  G.G. Krylov,  E.M. Ovsiyuk, V.M. Red'kov.
Maxwell electrodynamics in complex form, solutions with cylindric  symmetry
in the Riemann space. Doklady Natsionalnoi Akademii Nauk Belarusi.  {\bf 33},   52 -- 58 (2009).


\bibitem{6}
A.A. Bogush, V.M. Red'kov, G.G. Krylov.
Quantum-mechanical particle in a uniform magnetic field in spherical space $S_{3}$.
 Proceedings of the National Academy of Sciences of Belarus. Ser. fiz.-mat. {\bf 2}, 57 -- 63 (2009).




\bibitem{7}
 V.V. Kudriashov, Yu.A. Kurochkin, E.M. Ovsiyuk, V.M.
Red'kov. Motion caused by magnetic field in Lobachevsky space. AIP
Conference Proceedings. Vol. 1205, P. 120 -- 126 (2010); Eds. Remo
Ruffini and Gregory Vereshchagin. The sun, the stars, the Uiverse
and General relativity. International Conference in Honor of Ya.B. Zeldovich.
April 20-23, 2009, Minsk.



\bibitem{8}
V.V. Kudryashov, Yu.A. Kurochkin, E.M. Ovsiyuk, V.M.  Red'kov.
Motion of a particle in magnetic field in the Lobachevsky space.
Doklady Natsionalnoi Akademii Nauk Belarusi.  {\bf 53},  50--53 (2009).


\bibitem{9}
V.V. Kudryashov, Yu.A. Kurochkin, E.M. Ovsiyuk, V.M.
Red'kov. Classical Particle in Presence of Magnetic Field,
Hyperbolic Lobachevsky and Spherical Riemann Models.
 SIGMA {\bf  6}, 004, 34 pages (2010).

\bibitem{10}
V.M.  Red'kov.
Fields in Riemannian space  and the Lorentz group.  Publishing House "Belarusian Science",
Minsk  (2009)

\end{thebibliography}
\end{document}